\documentclass[twocolumn,showpacs,preprintnumbers,pra,hyperref]{revtex4}
\usepackage{amssymb}
\usepackage{amsfonts}
\usepackage{amsmath}
\usepackage{graphicx}

\begin{document}

\title{New Quantum Theory of Laser Cooling}
\author{Xiang-Yao Wu$^{a}$\thanks{%
E-mail: wuxy2066@163.com}, Xiao-Jing Liu$^{a}$, Bai-Jun
Zhang$^{a}$, Nuo Ba$^{a}$,  Yi-Heng Wu$^{a}$, Qing-Cai Wang$^{a}$,
Yan Wang$^{a}$, Nuo Ba$^{a}$, and Guang-Huai Wang$^{a}$}
\affiliation{$^{a}${\small Institute of Physics, Jilin Normal
University, Siping 136000, China}}

\begin{abstract}

{\small In this paper, we study the laser cooling mechanisms with
new Schrodinger quantum wave equation, which can describe a
particle in conservative and non-conservative force field. We
prove the atom in laser field can be cooled with the new theory,
and predict that the atom cooling temperature $T$ is directly
proportional to the atom vibration frequency $\omega$, which is in
accordance with experiment result.\newline \textbf{PACS:}
 03.65.-w, 37.10.De,  37.10.Mn \newline \textbf{Keywords:} Quantum theory; Atom cooling; cooling temperature}
\end{abstract}

\maketitle

\section{Introduction}

During the last decade, significant progress has been achieved in
laser cooling of lanthanides. Laser-cooled lanthanides are
effectively used in such fundamental fields as the study of cold
collisions \cite{s1}, Bose-Einstein condensation \cite{s2},
ultra-precise atomic clocks \cite{s3} and also open new
perspectives for implementation in nano-technology \cite{s4} and
quantum information \cite{s5}. In contrast to recently
demonstrated method of buffer gas cooling and trapping of
lanthanides in a magnetic dipole trap \cite{s6}, high precision
spectroscopy \cite{s7}, atomic frequency standards
\cite{s8,s9,s10}, Bose-Einstein condensation \cite{s11}, atomic
nanolitography \cite{s12,s13} and so on.

Today, laser cooling of atoms is widely used in experiments where
high precision spectroscopy or precise control of the atomic
motion is necessary. A large variety of schemes have been proposed
and applied to suit atoms with specific level structures and for
obtaining particular temperature ranges \cite{s14}. In general,
the lower the wanted temperature, the more sensitive does the
light scattering process has to be on the velocity of the atom.
For the most simple laser cooling scheme relying on the Doppler
shift of an optical transition, Doppler laser cooling, this means
the narrower the line width of the optical transition, the lower
the obtainable temperature. However, since the maximum cooling
force in the Doppler cooling scheme is dependent on the photon
scattering rate, and narrow line width transitions will lead to
longer cooling time than wider transitions.

In the theory of Laser cooling, there are semiclassical method for
Sisyphus cooling \cite{s15}, and showed that this method gives
excellent agreement with the fully quantum- mechanical method
\cite{s16}. In the semiclassical method the external degrees of
freedom, i.e., position and momentum, are treated as
simultaneously well-defined classical variables. The internal
degree of freedom, i.e., the magnetic substate, is treated fully
quantum mechanically, allowing for arbitrary superpositions.

On the other hand, various nano-mechanical resonators have been
investigated \cite{s17} extensively in recent years. To reveal the
quantum effect in the nano-mechanical devices, various cooling
schemes \cite{s18,s19,s20,s21,s22,s23} were proposed to drive them
to reach the standard quantum limit \cite{s24}. A famous one among
them is the optical radiation-pressure cooling scheme \cite{s12}
attributed to the sideband cooling \cite{s20,s21,s22,s23}, which
was previously well-developed to cool the spatial motion of the
trapped ions \cite{s25} or the neutral atoms \cite{s26}.

In this paper, we study the laser cooling mechanisms with new
Schrodinger quantum wave equation, which can describe the particle
in conservative and non-conservative force field \cite{s27}. We
prove the atom can be cooled in laser field, i,e, the atom
velocity approach zero, and find the cooling temperature of atom
$T$ is directly proportional to the atom vibration frequency
$\omega$, which is in accordance with experiment result.

\section{ The radiation force of atom in light field}

A moving atom sees the light moves towards Doppler shifted closer
to resonance, whereas the light is shifted away from resonance.
Thus, the atom predominantly scatters photons from the forward
direction and is slowed down. As the Doppler effect plays a
central role, the process is normally referred to as Doppler
cooling. Although the cooling process is quantum mechanical in
nature, as represented by the discrete momentum steps, the atomic
motion may be treated classically, if the atomic wave-packet is
well-localised in position and momentum space. In this case, the
time-averaged interaction can be separated into a mean cooling
force, and a diffusive term which accounts for the stochastic
nature of the spontaneous emission. For Doppler cooling, the
cooling force is generally obtained by treating the two beams
independently. The extension of Doppler cooling to three
dimensions is obvious. By using six beams, forming three
orthogonal standing waves, an atom will everywhere see a viscous
force.

Ring-like spatial distributions (modes) of atoms orbiting around a
core were firstly observed in a misaligned cesium MOT [28] and
explained in terms of the conventional MOT forces acting on each
individual atom plus the assumption about influence of the
collective interatomic forces acting between the trapped atoms
[28,29]. After observation in sodium MOT, the variety of spatial
structures of cooled atoms, a simple model of coordinate-dependent
vortex forces was developed which allowed to explain all observed
cooled atoms structures and the transitions between them in terms
of forces acting on each individual atom [30]. Due to the
misalignment, the radiative force acting on the atom along the y
direction has an $x$ dependence and vice versa. In other words,
beside the velocity and field-intensity dependent terms in the
force expression an extra azimuthal component do appears, which is
referred to as the vortex force. It is clear from consideration of
the forces in $xy$ plane. For a Gaussian beam propagating exactly
along the x-direction, the velocity-independent part of the
radiative force has the form [31]

\begin{eqnarray}
\vec{F}=-k \vec{v}- \kappa \vec{r},
\end{eqnarray}
where $\vec{v}$ is atom velocity, $\vec{r}$ is atom position, $k$
is damp coefficient of atom in light field, $\kappa$ is elastic
recovery coefficient. The first term $-k \vec{v}$ is used for
laser cooling, which is a non-conservative force, and the second
term $-\kappa \vec{r}$ is used for laser trapping, which is a
conservative force corresponding potential energy $\frac{1}{2}
\kappa r^2$.

\section{ A New Quantum Theory of Laser Cooling}

We know that Schrodinger equation is only suitable for the
particle in conservative force field. For the particle in
non-conservative field, it is needed new quantum wave equation
describe it. Recently, we have proposed a new quantum wave
equation, which can describe the particle in conservative and
non-conservative force field [27]. It is

\begin{eqnarray} i\hbar\frac{\partial}{\partial
t}\Psi(\vec{r},t)=(-\frac{\hbar^{2}}{2m}\nabla^{2}+U(r)-i\hbar\frac{3k}{m})\Psi(\vec{r},t)
\end{eqnarray}
where $U(r)$ is potential energy, the term $-i\hbar\frac{3k}{m}$
corresponding non-conservative force $\vec{F}=-k\vec{v}$. The Eq.
(1) is the radiative force of atom in light field, which include
both conservative force and non-conservative force, and it can be
described by the equation
\begin{eqnarray}
i\hbar\frac{\partial}{\partial
t}\Psi(\vec{r},t)=(-\frac{\hbar^{2}}{2m}\nabla^{2}+\frac{1}{2}\kappa
r^{2}-i\hbar\frac{3k}{m})\Psi(\vec{r},t)
\end{eqnarray}
By the method of separation of variable
\begin{eqnarray}
\Psi(\vec{r},t)=\Psi(\vec{r})f(t),
\end{eqnarray}
the Eq. (2) becomes
\begin{eqnarray}
-\frac{\hbar^{2}}{2m}\nabla^{2}\Psi(\vec{r})+\frac{1}{2}\kappa
r^{2}\Psi(\vec{r})=(E+i\hbar\frac{3k}{m})\Psi(\vec{r}),
\end{eqnarray}
and
\begin{eqnarray}
f(t)=ce^{-\frac{i}{\hbar}Et}.
\end{eqnarray}
The wave function $\Psi(\vec{r})$ and energy $E$ can be written
plural form. let
\begin{equation}
\Psi(\vec{r})=R(\vec{r})+iS(\vec{r}),
\end{equation}
and
\begin{equation}
E=E_{1}+iE_{2},
\end{equation}
substituting Eqs. (7) and (8) into (5), we have
\begin{eqnarray}
&&-\frac{\hbar^{2}}{2m}\nabla^{2}R(\vec{r})-i\frac{\hbar^{2}}{2m}\nabla^{2}S(\vec{r})
+\frac{1}{2}\kappa  r^{2}R(\vec{r})+i\frac{1}{2}\kappa
r^{2}S(\vec{r}) \nonumber\\&=&
E_{1}R(\vec{r})-E_{2}S(\vec{r})-\hbar\frac{3k}{m}S(\vec{r})
\nonumber\\&&
+i(E_{1}S(\vec{r})+E_{2}R(\vec{r})+\hbar\frac{3k}{m}R(\vec{r})).
\end{eqnarray}
From Eq. (9), we can obtain
\begin{equation}
-\frac{\hbar^{2}}{2m}\nabla^{2}R(\vec{r})
=(E_{1}-\frac{1}{2}\kappa
r^{2})R(\vec{r})-(E_{2}+\hbar\frac{3k}{m})S(\vec{r}),
\end{equation}
and
\begin{equation}
-\frac{\hbar^{2}}{2m}\nabla^{2}S(\vec{r})
=(E_{1}-\frac{1}{2}\kappa
r^{2})S(\vec{r})+(E_{2}+\hbar\frac{3k}{m})R(\vec{r}),
\end{equation}
Eq. (10) and (11) are multiplied by $R(\vec{r})$ and $S(\vec{r})$
respectively, we have
\begin{eqnarray}
&&-\frac{\hbar^{2}}{2m}\nabla^{2}R(\vec{r})\cdot R(\vec{r})
\nonumber\\&=&(E_{1}-\frac{1}{2}\kappa r^{2})R^{2}(\vec{r})
-(E_{2}+\hbar\frac{3k}{m})S(\vec{r})\cdot R(\vec{r}),
\end{eqnarray}
and
\begin{eqnarray}
&&-\frac{\hbar^{2}}{2m}\nabla^{2}S(\vec{r})\cdot S(\vec{r})
\nonumber\\&=&(E_{1}-\frac{1}{2}\kappa
r^{2})S^{2}(\vec{r})+(E_{2}+\hbar\frac{3k}{m})R(\vec{r})\cdot
S(\vec{r}),
\end{eqnarray}
the sum of Eq. (12) and (13) is
\begin{eqnarray}
&&-\frac{\hbar^{2}}{2m}\nabla^{2}R(\vec{r})\cdot
R(\vec{r})-\frac{\hbar^{2}}{2m}\nabla^{2}S(\vec{r})\cdot
S(\vec{r})\nonumber\\&=&(E_{1}-\frac{1}{2}\kappa
r^{2})R^{2}(\vec{r})+(E_{1}-\frac{1}{2}\kappa
r^{2})S^{2}(\vec{r}),
\end{eqnarray}
Eq. (14) can be written as
\begin{eqnarray}
-\frac{\hbar^{2}}{2m}\nabla^{2}R(\vec{r})\cdot R(\vec{r})
=(E_{1}-\frac{1}{2}\kappa  r^{2})R^{2}(\vec{r}),
\end{eqnarray}
and
\begin{eqnarray}
-\frac{\hbar^{2}}{2m}\nabla^{2}S(\vec{r})\cdot S(\vec{r})
=((E_{1}-\frac{1}{2}\kappa  r^{2})S^{2}(\vec{r}),
\end{eqnarray}
Eq. (10) and (11) are multiplied by $S(\vec{r})$ and $R(\vec{r})$
respectively, we have
\begin{eqnarray}
&&-\frac{\hbar^{2}}{2m}\nabla^{2}R(\vec{r})\cdot S(\vec{r})
\nonumber\\&=&(E_{1}-\frac{1}{2}\kappa  r^{2})R(\vec{r})\cdot
S(\vec{r})-(E_{2}+\hbar\frac{3k}{m})S^{2}(\vec{r}),
\end{eqnarray}
and
\begin{eqnarray}
&&-\frac{\hbar^{2}}{2m}\nabla^{2}S(\vec{r})\cdot R(\vec{r})
\nonumber\\&=&(E_{1}-\frac{1}{2}\kappa  r^{2})S(\vec{r})\cdot
R(\vec{r})+(E_{2}+\hbar\frac{3k}{m})R^{2}(\vec{r}),
\end{eqnarray}
the minus of Eq. (17) and (18) is
\begin{eqnarray}
&&-\frac{\hbar^{2}}{2m}\nabla^{2}R(\vec{r})\cdot
S(\vec{r})+\frac{\hbar^{2}}{2m}\nabla^{2}S(\vec{r})\cdot
R(\vec{r})\nonumber\\&=&-(E_{2}+\hbar\frac{3k}{m})(S^{2}(\vec{r})+R^{2}(\vec{r})),
\end{eqnarray}
and divided by $R(\vec{r})\cdot S(\vec{r})$ in Eq. (19), we have
\begin{eqnarray}
&&-\frac{\hbar^{2}}{2m}\frac{\nabla^{2}R(\vec{r})}{R(\vec{r})}+\frac{\hbar^{2}}{2m}\frac{\nabla^{2}S(\vec{r})}{S(\vec{r})}
\nonumber\\&=&-(E_{2}+\hbar\frac{3k}{m})\frac{S(\vec{r})}{R(\vec{r})}-(E_{2}+\hbar\frac{3k}{m})\frac{R(\vec{r})}{S(\vec{r})}.
\end{eqnarray}
From Eq. (15) and (16), we can find the left side of Eq. (20) is
zero, and Eq. (20) can be written as
\begin{eqnarray}
(E_{2}+\hbar\frac{3k}{m})(S^{2}(\vec{r})+R^{2}(\vec{r}))=0,
\end{eqnarray}
to get
\begin{eqnarray}
E_{2}=-\hbar\frac{3k}{m}.
\end{eqnarray}
In the following, we should solve Eqs. (15) and (16), they can be
written as
\begin{equation}
(-\frac{\hbar^{2}}{2m}\nabla^{2}+\frac{1}{2}\kappa
r^{2})R(\vec{r})=E_{1}R(\vec{r}),
\end{equation}
and
\begin{equation}
(-\frac{\hbar^{2}}{2m}\nabla^{2}+\frac{1}{2}\kappa
r^{2})S(\vec{r})=E_{1}S(\vec{r}),
\end{equation}
they are energy eigenequation of three-dimensional harmonic
oscillator. In rectangular coordinate system, The Eqs. (23) and
(24) eigenfunctions and eigenvalues are
\begin{equation}
R(\vec{r})=S(\vec{r})=\Psi_{n_{x}}(x)\Psi_{n_{y}}(y)\Psi_{n_{z}}(z),
\end{equation}
and
\begin{equation}
E_{1}=E_{N}=(N+\frac{3}{2})\hbar\omega, \hspace{0.3in}
N=0,1,2,3,\cdot\cdot\cdot
\end{equation}
where $\Psi_{n_{x}}(x)$, $\Psi_{n_{y}}(y)$ and $\Psi_{n_{z}}(z)$
are the wave functions of one-dimensional harmonic oscillator. The
Eq. (5) eigenfunction and eigenvalue is
\begin{eqnarray}
&&\Psi_{n_{x}n_{y}n_{z}}(x,y,z)\nonumber\\&=&\Psi_{n_{x}}(x)\Psi_{n_{y}}(y)\Psi_{n_{z}}(z)
+i\Psi_{n_{x}}(x)\Psi_{n_{y}}(y)\Psi_{n_{z}}(z),
\end{eqnarray}
and
\begin{equation}
E=E_{1}+iE_{2}=(N+\frac{3}{2})\hbar\omega-i\hbar\frac{3k}{m},
\end{equation}
the Eq. (3) particular solution is
\begin{eqnarray}
&&\Psi_{n_{x}n_{y}n_{z}}(x,y,z,t)=\Psi_{n_{x}n_{y}n_{z}}(x,y,z)e^{-\frac{i}{\hbar}Et}
\nonumber\\&=&\Psi_{n_{x}n_{y}n_{z}}(x,y,z)e^{-\frac{i}{\hbar}E_{1}t}\cdot
e^{-\frac{3k}{m}t}.
\end{eqnarray}
A atom velocity operator $\hat{v}$ is
\begin{equation}
\hat{v}=\frac{\hat{p}}{m}=\frac{\hbar}{m}\frac{1}{i}\nabla,
\end{equation}
at the state $\Psi_{n_{x}n_{y}n_{z}}(x,y,z,t)$, the expectation
value of velocity operator $\hat{v}$ is
\begin{equation}
\hat{v}(t)=\int\Psi^{*}_{n_{x}n_{y}n_{z}}(x,y,z,t)\hat{v}\Psi_{n_{x}n_{y}n_{z}}(x,y,z,t)d\vec{r},
\end{equation}
the expectation value of velocity component operator $\hat{v}_{x}$
is%
\begin{widetext}
\begin{eqnarray}
\bar{v}_{x}
&=&e^{-\frac{6k}{m}t}\int\Psi^{*}_{n_{x}n_{y}n_{z}}(x,y,z,t)\frac{\hbar}{m}\frac{1}{i}\frac{\partial}{\partial
x}\Psi_{n_{x}n_{y}n_{z}}(x,y,z,t)dxdydz\nonumber\\&=&
e^{-\frac{6k}{m}t}\frac{\hbar}{m}\frac{1}{i}\int{({\Psi}^{*}_{n_{x}}(x)
 \Psi^{*}_{n_{y}}(y)\Psi^{*}_{n_{z}}(z)-i\Psi^{*}_{n_{x}}(x)\Psi^{*}_{n_{y}}(y)
 \Psi^{*}_{n_{z}}(z)}
 \frac{\partial}{\partial
x}(\Psi_{n_{x}}(x)\Psi_{n_{y}}(y)\Psi_{{n_{z}}}(z)+i\Psi_{n_{x}}(x)\Psi_{n_{y}}(y)\Psi_{n_{z}}(z))dx
dy dz
\nonumber\\&=&e^{-\frac{6k}{m}t}\frac{\hbar}{m}\frac{1}{i}\int{(1-i)\Psi^{*}_{n_{x}}(x)
\Psi^{*}_{n_{y}}(y)\Psi^{*}_{n_{z}}(z)(1+i)\frac{\partial}{\partial
x}\Psi_{n_{x}}(x)\Psi_{n_{y}}(y)\Psi_{n_{z}}(z)}dx dy dz
\nonumber\\&=&e^{-\frac{6k}{m} t} \frac{\hbar}{m}
\frac{2\alpha}{i}\int\Psi^{*}_{n_{x}}(x)\Psi^{*}_{n_{y}}(y)\Psi^{*}_{n_{z}}(z)\cdot
(\sqrt{\frac{n_{x}}{2}}\Psi_{n_{x}-1}(x)-\sqrt{\frac{n_{x}+1}{2}}\Psi_{n_{x}+1}(x))
 \Psi_{n_{y}}(y)\Psi_{n_{z}}(z)dx dy dz
\nonumber\\
&=&e^{-\frac{6k}{m}t}\frac{\hbar}{m}\frac{\alpha}{i}\int\Psi^{*}_{n_{x}}
(\sqrt{\frac{n_{x}}{2}}\Psi_{n_{x}-1}(x)-\sqrt{\frac{n_{x}+1}{2}}\Psi_{n_{x}+1}(x))dx
\int\Psi^{*}_{n_{y}}(y)\Psi_{n_{y}}(y)dy\int\Psi^{*}_{n_{z}}(z)\Psi_{n_{z}}(z)dz
=0, \label{13}
\end{eqnarray}
\end{widetext}
with $\alpha=\sqrt{\frac{m\omega}{\hbar}}$, similarly, there are
\begin{eqnarray}
\bar{v}_{y}=0,  \hspace{0.3in} \bar{v}_{z}=0,
\end{eqnarray}%
the expectation value of velocity square component operator
$\hat{v}_{x}^{2}$ is
\begin{widetext}
\begin{eqnarray}
\overline{v_{x}^{2}}
&&=-\int\Psi^{*}_{n_{x}n_{y}n_{z}}(x,y,z,t)\frac{\hbar^{2}}{m^{2}}\frac{\partial^{2}}{\partial
x^{2}}\Psi_{n_{x}n_{y}n_{z}}(x,y,z,t)dxdydz\nonumber\\&&=-
e^{-\frac{6k}{m}t}\frac{\hbar^{2}}{m^{2}}\int(\Psi^{*}_{n_{x}}(x)
 \Psi^{*}_{n_{y}}(y)\Psi^{*}_{n_{z}}(z)-i\Psi^{*}_{n_{x}}(x)\Psi^{*}_{n_{y}}(y)
 \Psi^{*}_{n_{z}}(z))
\nonumber\\&&\hspace{0.9in}
 \frac{\partial^{2}}{\partial
x^{2}}(\Psi_{n_{x}}(x)\Psi_{n_{y}}(y)\Psi_{{n_{z}}}(z)+i\Psi_{n_{x}}(x)\Psi_{n_{y}}(y)\Psi_{n_{z}}(z))dxdydz
\nonumber\\&&=-
e^{-\frac{6k}{m}t}\frac{\hbar^{2}}{m^{2}}\int(1-i)\Psi^{*}_{n_{x}}(x)
\Psi^{*}_{n_{y}}(y)\Psi^{*}_{n_{z}}(z)\frac{\partial^{2}}{\partial
x^{2}}(1+i)\Psi_{n_{x}}(x) \Psi_{n_{y}}(y)\Psi_{n_{z}}(z)dxdydz
\nonumber\\&&=-e^{-\frac{6k}{m}t}\frac{2\hbar^{2}}{m^{2}}
\int\Psi^{*}_{n_{x}}(x)\Psi^{*}_{n_{y}}(y)
\Psi^{*}_{n_{z}}(z)\frac{\alpha^{2}}{2}(\sqrt{n_{x}(n_{x}-1)}
\Psi_{n_{x}-2}(x)-(2n_{x}+1)\Psi_{n_{x}}(x) \nonumber\\&&
\hspace{0.9in}+\sqrt{(n_{x}+1)(n_{x}+2)}\Psi_{n_{x}+2}(x))\Psi_{n_{y}}(y)
\Psi_{n_{z}}(z)dxdydz \nonumber\\&&=e^{-\frac{6k}{m}t}
{\alpha^{2}} \frac{\hbar^{2}}{m^{2}}
\int\Psi^{*}_{n_{x}}(x)\Psi^{*}_{n_{y}}(y)
\Psi^{*}_{n_{z}}(z)(2n_{x}+1)\Psi_{n_{x}}(x)\Psi_{n_{y}}(y)
\Psi_{n_{z}}(z)dxdydz \nonumber\\&&= e^{-\frac{6k}{m}t}
{\alpha^{2}} \frac{\hbar^{2}}{m^{2}}
\int(2n_{x}+1)\Psi^{*}_{n_{x}}(x)\Psi_{n_{x}}(x)dx\int\Psi^{*}_{n_{y}}(y)\Psi_{n_{y}}(y)dy
\int\Psi^{*}_{n_{z}}(z)\Psi_{n_{z}}(z)dz \nonumber\\
&&=e^{-\frac{6k}{m}t}\frac{\hbar^{2}}{m^{2}}\frac{m\omega}{\hbar}(2n_{x}+1)
=e^{-\frac{6k}{m}t}\frac{\hbar\omega}{m}(2n_{x}+1),
\end{eqnarray}
\end{widetext}
similarly, there are
\begin{eqnarray}
\overline{v_{y}^{2}}=e^{-\frac{6k}{m}t}\frac{\hbar\omega}{m}(2n_{y}+1),
\end{eqnarray}
and
\begin{eqnarray}
\overline{v_{z}^{2}}=e^{-\frac{6k}{m}t}\frac{\hbar\omega}{m}(2n_{z}+1),
\end{eqnarray}
From Eqs. (34)-(36), we can find when time $t$ increases
$\overline{v_{x}^{2}}\rightarrow 0$,
$\overline{v_{y}^{2}}\rightarrow 0$ and
$\overline{v_{z}^{2}}\rightarrow 0$, i.e., as time increases the
atom in the laser field (particular solution (29)) should be
cooled. In the following, we prove the atom can be cooled in
general solution, and the general solution is
\begin{eqnarray}
&&\Psi(x,y,z,t)=\sum_{n_{x}n_{y}n_{z}}C_{n_{x}n_{y}n_{z}}\Psi_{n_{x}n_{y}n_{z}}(x,y,z,t)
\nonumber\\&=&\sum_{n_{x}n_{y}n_{z}}C_{n_{x}n_{y}n_{z}}(1+i)\Psi_{n_{x}}(x)
\Psi_{n_{y}}(y)\Psi_{n_{z}}(z)\nonumber\\&&e^{-\frac{i}{\hbar}E_{1}t}e^{-\frac{3k}{m}t},
\end{eqnarray}
and the complex conjugate of the general solution is
\begin{widetext}
\begin{eqnarray}
\Psi^{*}(x,y,z,t)=\sum_{n'_{x}n'_{y}n'_{z}}C^*_{n'_{x}n'_{y}n'_{z}}(1-i)
\Psi^{*}_{n'_{x}}(x)\Psi^{*}_{n'_{y}}(y)\Psi^{*}_{n'_{z}}(z)e^{-\frac{i}{\hbar}E'_{1}t}e^{-\frac{3k}{m}t},
\end{eqnarray}
where $C_{n_{x}n_{y}n_{z}}$ and $C^*_{n'_{x}n'_{y}n'_{z}}$ are
superposition  coefficients, and $E_{1}$ and $E'_{1}$ are energy
levels, they are
\begin{eqnarray}
E_{1}=(n_{x}+n_{y}+n_{z}+\frac{3}{2})\hbar \omega,
\hspace{0.3in}n_{x}, n_{y}, n_{z}=0,1,2,3,\cdot\cdot\cdot
\end{eqnarray}
and
\begin{eqnarray}
E'_{1}=(n'_{x}+n'_{y}+n'_{z}+\frac{3}{2})\hbar \omega,
\hspace{0.3in}n'_{x}, n'_{y}, n'_{z}=0,1,2,3,\cdot\cdot\cdot
\end{eqnarray}
the expectation value of velocity component operator $\hat{v}_{x}$
is
\begin{eqnarray}
\bar{v}_{x}&=&e^{-\frac{6k}{m}t}\int\sum_{n'_{x}n'_{y}n'_{z}}\sum_{n_{x}n_{y}n_{z}}
C^*_{n'_{x}n'_{y}n'_{z}}C_{n_{x}n_{y}n_{z}}
(1-i)(1+i)\Psi^{*}_{n'_{x}}(x)
\Psi^{*}_{n'_{y}}(y)\Psi^{*}_{n'_{z}}(z)\nonumber\\&&
\hspace{0.5in} \frac{\hbar}{m}\frac{1}{i}\frac{\partial}{\partial
x}\Psi_{n_{x}}(x)
\Psi_{n_{y}}(y)\Psi_{n_{z}}(z)e^{-\frac{i}{\hbar}(E_{1}-E^{'}_{1})t}dxdydz
\nonumber\\&=&\frac{2\hbar}{m}\frac{1}{i}e^{-\frac{6k}{m}t}\sum_{n'_{x}n'_{y}n'_{z}}\sum_{n_{x}n_{y}n_{z}}
C^*_{n'_{x}n'_{y}n'_{z}}C_{n_{x}n_{y}n_{z}}\int\Psi^{*}_{n'_{x}}(x)
\Psi^{*}_{n'_{y}}(y)\Psi^{*}_{n'_{z}}(z)\nonumber\\&&\hspace{0.3in}\alpha(\sqrt{\frac{n_{x}}{2}}
\Psi_{n_{x}-1}(x)-\sqrt{\frac{n_{x}+1}{2}}\Psi_{n_{x}+1}(x))
\Psi_{n_{y}}(y)\Psi_{n_{z}}(z)e^{-\frac{i}{\hbar}(E_{1}-E^{'}_{1})t}dxdydz
\nonumber\\&=&\frac{2\hbar}{m}\frac{\alpha}{i}e^{-\frac{6k}{m}t}\sum_{n'_{x}n'_{y}n'_{z}}\sum_{n_{x}n_{y}n_{z}}
C^*_{n'_{x}n'_{y}n'_{z}}C_{n_{x}n_{y}n_{z}}\int\Psi^{*}_{n'_{x}}(x)
(\sqrt{\frac{n_{x}}{2}}
\Psi_{n_{x}-1}(x)-\sqrt{\frac{n_{x}+1}{2}}\Psi_{n_{x}+1}(x))dx\nonumber\\&&\int\Psi^{*}_{n'_{y}}(y)
\Psi_{n_{y}}(y)dy \Psi^{*}_{n'_{z}}(z)\Psi_{n_{z}}(z)dz
e^{-\frac{i}{\hbar}(E_{1}-E'_{1})t}
\nonumber\\&=&\frac{2\hbar}{m}\frac{\alpha}{i}e^{-\frac{6k}{m}t}\sum_{n'_{x}n'_{y}n'_{z}}\sum_{n_{x}n_{y}n_{z}}
C^*_{n'_{x}n'_{y}n'_{z}}C_{n_{x}n_{y}n_{z}} \nonumber\\
&&(\sqrt{\frac{n_{x}}{2}}\delta_{n'_{x},n_{x}-1 }-
\sqrt{\frac{n_{x}+1}{2}}\delta_{n'_{x}, n_{x}+1
})\delta_{n'_{y},n_{y}}\delta_{n'_{z},
n_{z}}e^{-\frac{i}{\hbar}(E_{1}-E^{'}_{1})t}
\nonumber\\&=&\frac{2\hbar}{m} \frac{\alpha}{i}
e^{-\frac{6k}{m}t}\sum_{n_{x}n_{y}n_{z}}( \sqrt{\frac{n_{x}}{2}}
C_{n_{x}n_{y}n_{z}} C^*_{n_{x}-1 n_{y}n_{z}}
e^{-\frac{i}{\hbar}(n_{x}-n_{x}+1)\hbar\omega  t} \nonumber\\&&-
\sqrt{\frac{n_{x}+1}{2}} C_{n_{x}n_{y}n_{z}} C^*_{n_{x}+1
n_{y}n_{z}} e^{-\frac{i}{\hbar}(n_{x}-n_{x}-1)\hbar\omega t})
\nonumber\\&=&\frac{2\hbar}{m} \frac{\alpha}{i}
e^{-\frac{6k}{m}t}\sum_{n_{x}n_{y}n_{z}}( \sqrt{\frac{n_{x}}{2}}
C_{n_{x}n_{y}n_{z}}C^*_{n_{x}-1 n_{y}n_{z}}  e^{-i\omega t}
-\sqrt{\frac{n_{x}+1}{2}}C_{n_{x}n_{y}n_{z}} C^*_{n_{x}+1
n_{y}n_{z}} e^{i\omega t}),
\end{eqnarray}
the expectation value of velocity square component operator
$\hat{v}_{x}^{2}$ is
\begin{eqnarray}
\overline{v_{x}^{2}}
&=&\int\Psi^{*}(x,y,z,t)(-\frac{\hbar^{2}}{m^{2}}\frac{\partial^{2}}{\partial
x^{2}})\Psi(x,y,z,t)dxdydz\nonumber\\&=&
(-\frac{\hbar^{2}}{m^{2}})e^{-\frac{6k}{m}t}\int\sum_{n'_{x}n'_{y}n'_{z}}\sum_{n_{x}n_{y}n_{z}}
C^*_{n'_{x}n'_{y}n'_{z}} C_{n_{x}n_{y}n_{z}}
\Psi^{*}_{n'_{x}}(x)\Psi^{*}_{n'_{y}}(y)\Psi^{*}_{n'_{z}}(z)
\nonumber\\&&(1-i)(1+i)\frac{\partial^{2}}{\partial
x^{2}}(\Psi_{n_{x}}(x)
 \Psi_{n_{y}}(y)\Psi_{n_{z}}(z)) e^{-\frac{i}{\hbar}(E_{1}-E'_{1})t}dxdydz
\nonumber\\&=&-\frac{2\hbar^{2}}{m^{2}}e^{-\frac{6k}{m}t}\sum_{n'_{x}n'_{y}n'_{z}}\sum_{n_{x}n_{y}n_{z}}
C^*_{n'_{x}n'_{y}n'_{z}}
C_{n_{x}n_{y}n_{z}}\int\Psi^{*}_{n'_{x}}(x)
 \Psi^{*}_{n'_{y}}(y)\Psi^{*}_{n'_{z}}(z)\frac{\alpha^{2}}{2}
[\sqrt{n_{x}(n_{x}-1)}\Psi_{n_{x}-2}(x)\nonumber\\ &&-(2n_{x}+1)
\Psi_{n_{x}}(x)+\sqrt{(n_{x}+1)(n_{x}+2)}\Psi_{n_{x}+2}(x)]\Psi_{n_{y}}(y)
\Psi_{n_{z}}(z)dxdydz
\nonumber\\&=&-\frac{\hbar^{2}\alpha^{2}}{m^{2}}e^{-\frac{6k}{m}t}
\sum_{n'_{x}n'_{y}n'_{z}}\sum_{n_{x}n_{y}n_{z}}
C^*_{n'_{x}n'_{y}n'_{z}}
C_{n_{x}n_{y}n_{z}}[\sqrt{n_{x}(n_{x}-1)}\delta_{n'_{x},n_{x}-2}-(2n_{x}+1)\delta_{n'_{x},n_{x}}
\nonumber\\&&+\sqrt{(n_{x}+1)(n_{x}+2)}\delta_{n'_{x},
n_{x}+2}]\delta_{n'_{y},n_{y}}\delta_{n'_{z},n_{z}}
e^{-\frac{i}{\hbar}(E_{1}-E'_{1})t} \nonumber\\
&=&-\frac{\hbar^{2}\alpha^{2}}{m^{2}}e^{-\frac{6k}{m}t}
\sum_{n_{x}n_{y}n_{z}}[\sqrt{n_{x}(n_{x}+1)} C_{n_{x}n_{y}n_{z}}
C^*_{n_{x}-2n_{y}n_{z}}e^{-\frac{i}{\hbar}(n_{x}-n_{x}+2)\hbar\omega
t}\nonumber\\&&-(2n_{x}+1) C_{n_{x}n_{y}n_{z}}
C^*_{n_{x}n_{y}n_{z}}e^{-\frac{i}{\hbar}(n_{x}-n_{x})\hbar\omega
t} \nonumber\\&&+\sqrt{(n_{x}+1)(n_{x}+2)}C_{n_{x}n_{y}n_{z}}
C^*_{n_{x}+2n_{y}n_{z}}e^{-\frac{i}{\hbar}(n_{x}-n_{x}-2)\hbar\omega
t}] \nonumber
\end{eqnarray}
\begin{eqnarray}
&=&-\frac{\hbar \omega}{m}e^{-\frac{6k}{m}t}
\sum_{n_{x}n_{y}n_{z}}[\sqrt{n_{x}(n_{x}+1)} C_{n_{x}n_{y}n_{z}}
C^*_{n_{x}-2n_{y}n_{z}}e^{-2i\omega t}\nonumber\\&&-(2n_{x}+1)
C_{n_{x}n_{y}n_{z}}
C^*_{n_{x}n_{y}n_{z}}+\sqrt{(n_{x}+1)(n_{x}+2)}C_{n_{x}n_{y}n_{z}}
C^*_{n_{x}+2n_{y}n_{z}}e^{2i\omega t}],
\end{eqnarray}
In Eqs. (41) and (42), the series
\begin{eqnarray}
\sum_{n_{x}n_{y}n_{z}}( \sqrt{\frac{n_{x}}{2}}
C_{n_{x}n_{y}n_{z}}C^*_{n_{x}-1 n_{y}n_{z}} e^{-i\omega t}
-\sqrt{\frac{n_{x}+1}{2}}C_{n_{x}n_{y}n_{z}} C^*_{n_{x}+1
n_{y}n_{z}} e^{i\omega t}),
\end{eqnarray}
 and
\begin{eqnarray}
&&\sum_{n_{x}n_{y}n_{z}}(\sqrt{n_{x}(n_{x}+1)} C_{n_{x}n_{y}n_{z}}
C^*_{n_{x}-2n_{y}n_{z}}e^{-2i\omega t}-(2n_{x}+1)
C_{n_{x}n_{y}n_{z}}
C^*_{n_{x}n_{y}n_{z}}\nonumber\\&&\hspace{0.5in}+\sqrt{(n_{x}+1)(n_{x}+2)}C_{n_{x}n_{y}n_{z}}
C^*_{n_{x}+2n_{y}n_{z}}e^{2i\omega t}),
\end{eqnarray}
are convergent, when time $t$ increases $\hat{v}_{x}\rightarrow 0$
and $\hat{v}_{x}^{2}\rightarrow 0$ ($\hat{v}_{y}\rightarrow 0$,
$\hat{v}_{y}^{2}\rightarrow 0$ and $\hat{v}_{z}\rightarrow 0$,
$\hat{v}_{z}^{2}\rightarrow 0$), i.e., as time increases the atom
in the laser field (general solution (37)) should be cooled.\\
For three-dimensional harmonic oscillator, the wave functions are
degenerate, and the degeneracy is
\begin{eqnarray}
f=\frac{1}{2}(N+1)(N+2),  \hspace{0.3in}N=0,1,2,3\cdots
\end{eqnarray}

the quantum number $N$ and corresponding wave equation $\Psi_{N}$
are

\hspace{0.4in} $N=0,\psi_{000}$,

\hspace{0.4in} $N=1,\psi_{100},\psi_{010},\psi_{001}$,

\hspace{0.4in}
$N=2,\psi_{110},\psi_{101},\psi_{011},\psi_{200},\psi_{002}$,

\hspace{0.4in}
$N=3,\psi_{111},\psi_{102},\psi_{120},\psi_{210},\psi_{021},\psi_{012},\psi_{201},\psi_{300},
  \psi_{030},\psi_{003}$,

\hspace{0.4in}
$N=4,\psi_{112},\psi_{121},\psi_{211},\psi_{301},\psi_{031},\psi_{103},\psi_{130},\psi_{301},
  \psi_{013},\psi_{004},\psi_{040},\psi_{400},\psi_{202},\psi_{022},\psi_{220}$,

\hspace{0.4in} $\cdots$

the total wave function can be written as:
\begin{eqnarray}
\psi(x,y,z,t)&=&(1+i)[C_{0}\psi_{0}(x,y,z)e^{-\frac{i}{\hbar}E_{0}t}e^{-\frac{3k}{m}t}
+C_{1}\psi_{1}(x,y,z)e^{-\frac{i}{\hbar}E_{1}t}e^{-\frac{3k}{m}t}
\nonumber\\&&+C_{2}\psi_{2}(x,y,z)e^{-\frac{i}{\hbar}E_{2}t}e^{-\frac{3k}{m}t}
+\cdots
+C_{N}\psi_{N}(x,y,z)e^{-\frac{i}{\hbar}E_{N}t}e^{-\frac{3k}{m}t}
+\cdots]
\nonumber\\&=&(1+i)e^{-\frac{3k}{m}t}[C_{000}\psi_{0}(x)\psi_{0}(y)\psi_{0}(z)e^{-i\frac{3}
 {2}\omega t}+C_{100}\psi_{1}(x)\psi_{0}(y)\psi_{0}(z)e^{-i\frac{5}
 {2}\omega t}\nonumber\\&&+C_{010}\psi_{0}(x)\psi_{1}(y)\psi_{0}(z)e^{-i\frac{5}
 {2}\omega t}+C_{001}\psi_{0}(x)\psi_{0}(y)\psi_{1}(z)e^{-i\frac{5}
 {2}\omega t}\nonumber\\&&+C_{110}\psi_{1}(x)\psi_{1}(y)\psi_{0}(z)e^{-i\frac{7}
 {2}\omega t}+C_{101}\psi_{1}(x)\psi_{0}(y)\psi_{1}(z)e^{-i\frac{7} {2}\omega t}
+\cdots].
\end{eqnarray}
The real measurement value of $\hat{v}_{x}^{2}$ is its average
value in a period. It is
\begin{eqnarray}
\langle\overline{v_{x}^{2}}\rangle&=&\frac{1}{T}\int^{T}_{0}\overline{v_{x}^{2}}dt
\nonumber\\&=&\frac{\hbar\omega}{m}e^{-\frac{6k}{m}t}
\sum_{n_{x}n_{y}n_{z}}(2n_{x}+1)C_{n_{x}n_{y}n_{z}}C^{*}_{n_{x}n_{y}n_{z}},
\end{eqnarray}
from Eq. (47), we have
\begin{eqnarray}
\langle\overline{v_{x}^{2}}\rangle&=&\frac{\hbar\omega}{m}e^{-\frac{6k}{m}t}
[|C_{000}|^{2}+3|C_{100}|^{2}+|C_{010}|^{2}+|C_{001}|^{2}
\nonumber\\&&+3|C_{110}|^{2}+3|C_{101}|^{2}|+|C_{011}|^{2}+5|C_{200}|^{2}+|C_{002}|^{2}+|C_{020}|^{2}
\nonumber\\&&+3|C_{111}|^{2}+3|C_{102}|^{2}+3|C_{120}|^{2}+5|C_{210}|^{2}+|C_{021}|^{2}+|C_{012}|^{2}
\nonumber\\&&+5|C_{201}|^{2}+7|C_{300}|^{2}+|C_{030}|^{2}+|C_{003}|^{2}
\nonumber\\&&+3|C_{112}|^{2}+3|C_{121}|^{2}+5|C_{211}|^{2}+7|C_{301}|^{2}+|C_{031}|^{2}
\nonumber\\&&+3|C_{103}|^{2}+3|C_{130}|^{2}+7|C_{301}|^{2}+|C_{013}|^{2}+|C_{004}|^{2}
\nonumber\\&&+|C_{040}|^{2}+9|C_{400}|^{2}+5|C_{202}|^{2}+|C_{022}|^{2}+5|C_{220}|^{2}+\cdots]
\nonumber\\&=&\frac{\hbar\omega}{m}e^{-\frac{6k}{m}t}
[|C_{000}|^{2}+5|C_{100}|^{2}+14|C_{110}|^{2}+30|C_{111}|^{2}+55|C_{112}|^{2}+\cdots].
\end{eqnarray}
\end{widetext}
According to Boltzmann distribution law, when the atom is at heat
balance state, the probability that atom is on
the energy level $E$ is directly proportional to $e^{-E/{k_{B}T}}$. \\

The probability of atom ground state is
\begin{equation}
Ce^{-\frac{E_{0}}{k_{B}T}},
\end{equation}
the probability of atom first excited state is
\begin{equation}
Ce^{-\frac{E_{1}}{k_{B}T}},
\end{equation}
the probability of atom N-th excited state is
\begin{equation}
Ce^{-\frac{E_{N}}{k_{B}T}},
\end{equation}
the total probability is equal to $1$, i.e.,
\begin{eqnarray}
Ce^{-\frac{E_{0}}{k_{B}T}}+Ce^{-\frac{E_{1}}{k_{B}T}}+\cdots
Ce^{-\frac{E_{N}}{k_{B}T}}+\cdots=1,
\end{eqnarray}
and
\begin{eqnarray}
C=\frac{1-e^{-\frac{\hbar\omega}{K_{B}T}}}{e^{-\frac{E_{0}}{K_{B}T}}}.
\end{eqnarray}
From Eq. (46), we can calculate the states probability. The ground
state probability is
\begin{eqnarray}
(1+i)^{2}|C_{000}|^{2}e^{-\frac{6k}{m}t}=Ce^{-\frac{E_{0}}{K_{B}T}}=1-e^{-\frac{\hbar\omega}{K_{B}T}},
\end{eqnarray}
and
\begin{eqnarray}
|C_{000}|^{2}e^{-\frac{6k}{m}t}=\frac{1-e^{-\frac{\hbar\omega}{K_{B}T}}}{2},
\end{eqnarray}
for the first excited state, there is
\begin{eqnarray}
(1+i)^{2}|C_{100}|^{2}e^{-\frac{6k}{m}t}&=&(1+i)^{2}|C_{010}|^{2}e^{-\frac{6k}{m}t}\nonumber\\&=&
(1+i)^{2}|C_{001}|^{2}e^{-\frac{6k}{m}t}
\nonumber\\&=&\frac{1}{3}Ce^{-\frac{E_{1}}{K_{B}T}}
\nonumber\\&=&\frac{1}{3}e^{-\frac{\hbar\omega}{K_{B}T}}(1-e^{-\frac{\hbar\omega}{K_{B}T}}),
\end{eqnarray}
and
\begin{eqnarray}
|C_{100}|^{2}e^{-\frac{6k}{m}t}&=&|C_{010}|^{2}e^{-\frac{6k}{m}t}=|C_{001}|^{2}e^{-\frac{6k}{m}t}\nonumber\\&=&
\frac{1}{6}e^{-\frac{\hbar\omega}{K_{B}T}}(1-e^{-\frac{\hbar\omega}{K_{B}T}}),
\end{eqnarray}
for the second excited state, there is
\begin{eqnarray}
2|C_{110}|^{2}e^{-\frac{6k}{m}t}=\frac{1}{6}Ce^{-\frac{E_{2}}{K_{B}T}},
\end{eqnarray}
and
\begin{eqnarray}
|C_{110}|^{2}e^{-\frac{6k}{m}t}
=\frac{1}{12}e^{-\frac{2\hbar\omega}{K_{B}T}}(1-e^{-\frac{\hbar\omega}{K_{B}T}}),
\end{eqnarray}
for the third excited state, there is
\begin{eqnarray}
2|C_{111}|^{2}e^{-\frac{6k}{m}t}=\frac{1}{10}Ce^{-\frac{E_{3}}{K_{B}T}},
\end{eqnarray}
and
\begin{eqnarray}
|C_{111}|^{2}e^{-\frac{6k}{m}t}
=\frac{1}{20}e^{-\frac{3\hbar\omega}{K_{B}T}}(1-e^{-\frac{\hbar\omega}{K_{B}T}}),
\end{eqnarray}
for the forth excited state, there is
\begin{eqnarray}
2|C_{112}|^{2}e^{-\frac{6k}{m}t}=\frac{1}{15}Ce^{-\frac{E_{4}}{K_{B}T}},
\end{eqnarray}
and
\begin{eqnarray}
|C_{112}|^{2}e^{-\frac{6k}{m}t}
=\frac{1}{30}e^{-\frac{4\hbar\omega}{K_{B}T}}(1-e^{-\frac{\hbar\omega}{K_{B}T}}),
\end{eqnarray}
substituting Eqs. (55), (57), (59), (61) and (63) into (48), we
have
\begin{widetext}
\begin{eqnarray}
\langle\overline{v_{x}^{2}}\rangle&=&\frac{\hbar\omega}{m}[\frac{1-e^{-\frac{\hbar\omega}{K_{B}T}}}{2}
+5\cdot\frac{1}{6}e^{-\frac{\hbar\omega}{K_{B}T}}(1-e^{-\frac{\hbar\omega}{K_{B}T}})
+14\cdot\frac{1}{12}e^{-\frac{2\hbar\omega}{K_{B}T}}(1-e^{-\frac{\hbar\omega}{K_{B}T}})
\nonumber\\&&+30\cdot\frac{1}{20}e^{-\frac{3\hbar\omega}{K_{B}T}}(1-e^{-\frac{\hbar\omega}{K_{B}T}})
+55\cdot\frac{1}{30}e^{-\frac{4\hbar\omega}{K_{B}T}}(1-e^{-\frac{\hbar\omega}{K_{B}T}})+\cdots]
\nonumber\\&=&\frac{\hbar\omega}{m}(1-e^{-\frac{\hbar\omega}{K_{B}T}})[\frac{1}{2}
+\frac{5}{6}e^{-\frac{\hbar\omega}{K_{B}T}}+\frac{7}{6}e^{-\frac{2\hbar\omega}{K_{B}T}}
+\frac{9}{6}e^{-\frac{3\hbar\omega}{K_{B}T}}+\frac{11}{6}e^{-\frac{4\hbar\omega}{K_{B}T}}+\cdots],
\end{eqnarray}
we define the series $S$
\begin{eqnarray}
S=\frac{1}{6}(5e^{-x}+7e^{-2x}
+9e^{-3x}+11e^{-4x}+13e^{-5x}+\cdots),
\end{eqnarray}
with $x=\frac{\hbar\omega}{K_{B}T}$, and the series $S_{0}$ is
\begin{eqnarray}
S_{0}=5e^{-x}+7e^{-2x} +9e^{-3x}+11e^{-4x}+13e^{-5x}+\cdots,
\end{eqnarray}
and
\begin{eqnarray}
S_{0}e^{-x}=5e^{-2x}+7e^{-3x}
+9e^{-4x}+11e^{-5x}+13e^{-6x}+\cdots,
\end{eqnarray}
so
\begin{eqnarray}
S_{0}-S_{0}e^{-x}&=&5e^{-x}+2e^{-2x}
+2e^{-3x}+2e^{-4x}+2e^{-5x}+2e^{-6x}+\cdots
\nonumber\\&=&5e^{-x}+2\cdot\lim_{N\rightarrow\infty}\frac{e^{-2x}-e^{(N+2)x}}{1-e^{-x}}
\nonumber\\&=&\frac{5e^{-x}-3e^{-2x}}{1-e^{-x}},
\end{eqnarray}
and
\begin{eqnarray}
S_{0}=\frac{5e^{-x}-3e^{-2x}}{(1-e^{-x})^{2}},
\end{eqnarray}
and so
\begin{eqnarray}
S=\frac{1}{6}S_{0}=\frac{1}{6}\frac{5e^{-x}-3e^{-2x}}{(1-e^{-x})^{2}},
\end{eqnarray}
substituting Eq. (70) into (64), we have
\begin{eqnarray}
\langle\overline{v_{x}^{2}}\rangle&=&\frac{\hbar\omega}{m}(1-e^{-\frac{\hbar\omega}{K_{B}T}})(\frac{1}{2}
+\frac{1}{6}\frac{5e^{-\frac{\hbar\omega}{K_{B}T}}-3e^{-\frac{2\hbar\omega}{K_{B}T}}}
{(1-e^{-\frac{\hbar\omega}{K_{B}T}})^{2}}).
\end{eqnarray}
The energy equipartition principle is
\begin{eqnarray}
\frac{1}{2}m \langle\overline{v_{x}^{2}}\rangle=\frac{1}{2}k_{B}T,
\end{eqnarray}
substituting Eq. (71) into (72), we have
\begin{eqnarray}
\frac{1}{2}\hbar\omega(1-e^{-\frac{\hbar\omega}{K_{B}T}})(\frac{1}{2}
+\frac{1}{6}\frac{5e^{-\frac{\hbar\omega}{k_{B}T}}-3e^{-\frac{2\hbar\omega}{K_{B}T}}}
{(1-e^{-\frac{\hbar\omega}{k_{B}T}})^{2}})=\frac{1}{2}k_{B}T,
\end{eqnarray}
i.e.,
\begin{eqnarray}
(1-e^{-\frac{\hbar\omega}{K_{B}T}})(\frac{1}{2}
+\frac{1}{6}\frac{5e^{-\frac{\hbar\omega}{k_{B}T}}-3e^{-\frac{2\hbar\omega}{K_{B}T}}}
{(1-e^{-\frac{\hbar\omega}{k_{B}T}})^{2}})=\frac{k_{B}T}{\hbar\omega}.
\end{eqnarray}
\end{widetext}
Eq. (74) is atom cooling temperature equation in laser field, we
can obtain the atom cooling temperature from the equation.

\section{ Numerical result}

Next, we present our numerical calculation of atom cooling
temperature. The Eq. (44) is transcendental equation, we can
obtain the cooling temperature by the following two functions
crossing point
\begin{eqnarray}
y_{1}=(1-e^{-\frac{\hbar\omega}{K_{B}T}})(\frac{1}{2}
+\frac{1}{6}\frac{5e^{-\frac{\hbar\omega}{k_{B}T}}-3e^{-\frac{2\hbar\omega}{K_{B}T}}}
{(1-e^{-\frac{\hbar\omega}{k_{B}T}})^{2}}),
\end{eqnarray}
\begin{eqnarray}
y_{2}=\frac{k_{B}T}{\hbar\omega}.
\end{eqnarray}

\begin{figure}[ptb]
\includegraphics[width=7 cm]{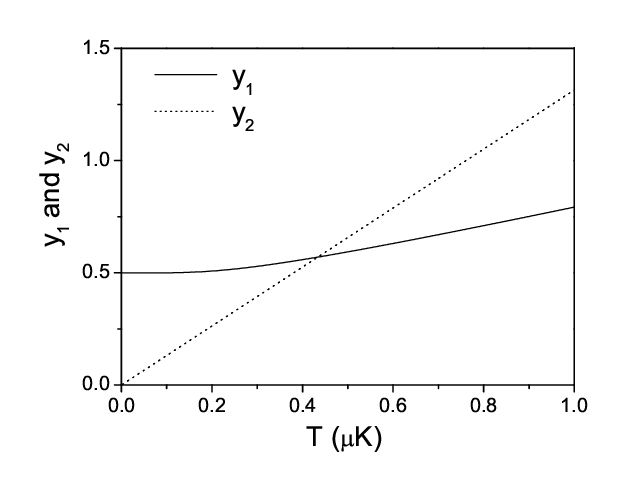}
\caption{The relation between $y_{1}$ , $y_{2}$ and temperature
$T$ when $\omega=100kHz$.} \label{Fig1}
\end{figure}

\begin{figure}[ptb]
\includegraphics[width=7 cm]{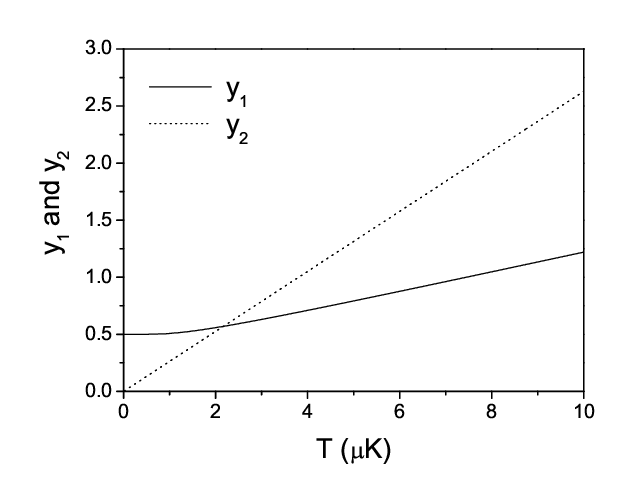}
\caption{The relation between $y_{1}$ , $y_{2}$ and temperature
$T$ when $\omega=500kHz$.} \label{Fig2}
\end{figure}

\begin{figure}[ptb]
\includegraphics[width=7 cm]{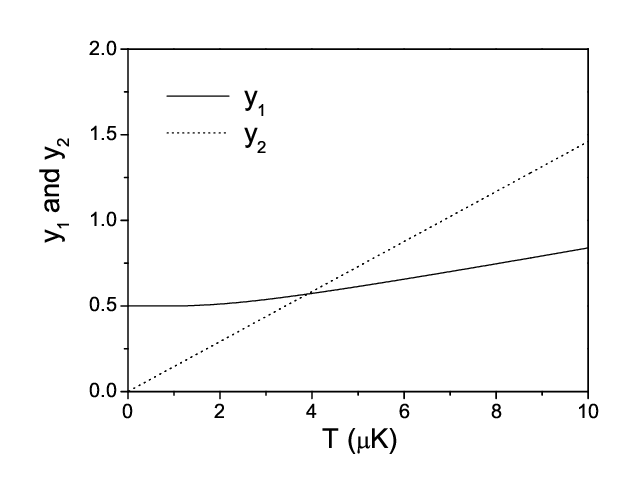}
\caption{The relation between $y_{1}$ , $y_{2}$ and temperature
$T$ when $\omega=900kHz$. } \label{Fig1}
\end{figure}

 The main input parameters are: Plank constant $\hbar=1.05\times
10^{-34} Js$, Boltzmann constant $k_{B}=1.38\times 10^{-23}
JK^{-1}$, in laser field, the atom vibration frequency $\omega$ is
about several hundred $kHz$. It exhibits that the functions
$y_{1}$, $y_{2}$ varies with the temperature $T$ in FIG. 1 to FIG.
3 corresponding to the different vibration frequencies $\omega$.
In FIG. 1, we take the vibration frequency $\omega=100kHz$, and
give the relation curve between functions $y_{1}$ , $y_{2}$ and
temperature $T$, and then obtain the atom cooling temperature
$T=0.433425\mu K$. In FIG. 2, we take the vibration frequency
$\omega=500kHz$, and give the relation curve between functions
$y_{1}$ , $y_{2}$ and temperature $T$, and then obtain the atom
cooling temperature $T=2.167125\mu K$. In FIG. 3, we take the
vibration frequency $\omega=900kHz$, and give the relation curve
between functions $y_{1}$ , $y_{2}$ and temperature $T$, and then
obtain the atom cooling temperature $T=3.90082\mu K$. From FIG. 1
to FIG. 3, we can also find that the atom cooling temperature $T$
gradually increase as the vibration frequency $\omega$ increase.
In FIG. 4, we give the relation between vibration frequency
$\omega$ and the atom cooling temperature $T$, and find the atom
cooling temperature is directly proportional to vibration
frequency. By calculation, we find the relation: $T=4.334\times
10^{-3}\omega$. In the formula, $T$ unit is $\mu K$, and $\omega$
unit is $KHz$. When the atom vibration frequency $\omega$ is in
the range of $100kHz\sim900kHz$, the atom cooling temperature $T$
is from $0.433\mu K$ to $3.901\mu K$. Recently, the authors A. D.
O'Connell and so on measure atom cooling temperature $T=25 mK$
when the atom vibration frequency $\omega=6.175 GHz$. By the
formula $T=4.334\times 10^{-3}\omega$, we obtain the atom cooling
temperature $T=26.76 mK$. They have also find atom corresponding
cooling temperature $T$ is directly proportional to the atom
vibration frequency $\omega$ [32]. These experiment results are in
accordance with our theory results. Obviously, by reducing the
atom vibration frequency $\omega$, we can achieve more lower atom
cooling temperature.

\begin{figure}[ptb]
\includegraphics[width=7 cm]{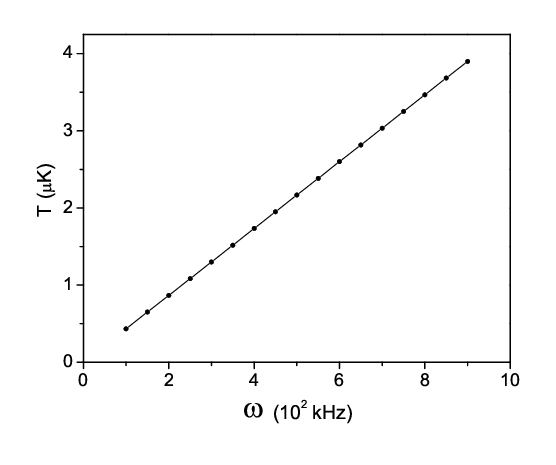}
\caption{The relation between temperature and frequency $\omega$.}
\label{Fig2}
\end{figure}

\section{ Conclusion}

We study the laser cooling mechanisms with new Schrodinger quantum
wave equation, which can describe the particle in conservative and
non-conservative force field. With the new theory, We prove the
atom can be cooled in laser field, and give the atom cooling
temperature in laser field. Otherwise, we give new prediction: (1)
the atom cooling temperature is directly proportional to the atom
vibration frequency. By calculation, we find they are:
$T=4.334\times 10^{-3}\omega$. (2) By reducing the atom vibration
frequency in laser field, we can achieve more lower atom cooling
temperature. these results are in accordance with recently
experiment results.


\newpage

\begin{thebibliography}{99}

\bibitem{s1}
 L. Santos, G.V. Shlyapnikov, P. Zoller and M. Lewenstein, Phys.
Rev. Lett. {\bf 85}, 1791 (2000).

\bibitem{s2}
 Y.Takasu, K.Maki, K.Komori, T.Takano, K.Honda, M.Kumakura,
T.Yabuzaki and Y.Takahashi, Phys. Rev. Lett. {\bf 91}, 040404
(2003).

\bibitem{s3}
 Z.W.Barber, C.W.Hoyt, C.W.Oates, L.Hollberg, A.V.Taichenachev,
and V.I.Yudin, Phys. Rev. Lett. {\bf 96}, 083002 (2006).

\bibitem{s4}
 S.B.Hill and J.J.McClelland, Appl. Phys. Lett. {\bf 82}, 3128 (2003).

\bibitem{s5}
 C. Monroe, Nature {\bf 416}, 238 (2002).

\bibitem{s6}
 C.I.Hancox, S.C.Doret, M.T. Hummon, L. Luo, J.M.Doyle, Nature
  {\bf 431}, 281 (2004).

\bibitem{s7}
J. W. R. Tabosa, S. S. Vianna and F. A. M. de Oliveira, Phys. Rev.
A {\bf 55}, 2968 (1997); T. M. Fortier, Y. Le Coq, J. E.
Stalnaker, D. Ortega, S. A. Diddams, C.W. Oates and L. Hollberg,
Phys. Rev. Lett. {\bf 97}, 163905 (2006).

\bibitem{s8}
D. J. Berkeland, J. D. Miller, J. C. Bergquist, W. M. Itano and D.
J. Wineland, Phys. Rev. Lett. {\bf 80}, 2089 (1998).

\bibitem{s9}
G. Santarelli, Ph. Laurent, P. Lemonde, A. Clairon, A. G. Mann, S.
Chang, A. N. Luiten and C. Salomon, Phys. Rev. Lett. {\bf 82},
4619 (1999).

\bibitem{s10}
C. Degenhardt, H. Stoehr, C. Lisdat, G. Wilpers, H. Schnatz, B.
Lip- phardt, T. Nazarova, P.-E. Pottie, U. Sterr, J. Helmcke and
F. Riehle, Phys. Rev. A {\bf 72}, 062111 (2005).

\bibitem{s11}
W.Ketterle, Usp.Fiz.Nauk, {\bf 173}, 12 (2003).

\bibitem{s12}
F. Lison, P. Schuh, D. Haubrich and D. Meschede Phys. Rev. A {\bf
61}, 013405 (2000).

\bibitem{s13}
D. V. Strekalov, A. Turlapov, A. Kumarakrishnan and T. Sleator
Phys. Rev. A {\bf 66}, 023601 (2002).

\bibitem{s14}
H. Metcalf and P. van der Straten, Laser cooling and trapping
(Springer, 1999).

\bibitem{s15}
S. Jonsell, C. M. Dion, M. Nylen, S. J. H. Petra, P. Sjolund, A.
Kastberg, Eur. Phys. J. D {\bf 39}, 3889 (2006).

\bibitem{s16}
J. Dalibard, Y. Castin, K. Momer, Phys. Rev. Lett. {\bf 68}, 580
(1992).

\bibitem{s17}
A. N. Clelandand and M. L. Roukes, Appl. Phys. Lett. {\bf 69},
2653 (1996); X. M. H. Huang et al., Nature {\bf 421}, 496 (2003).

\bibitem{s18}
I. Wilson-Rae et al., Phys. Rev. Lett. {\bf 92}, 075507 (2004); P.
Zhang et al., Phys. Rev. Lett. {\bf 95}, 097204 (2005); A. Naik et
al., Nature {\bf 443}, 193 (2006).

\bibitem{s19}
C. H. Metzger and K. Karrai, Nature (London) {\bf 432}, 1002
(2004); S. Gigan, et al., Nature (London) {\bf 444}, 67 (2006); O.
Arcizet et al., Nature (London) {\bf 444}, 71 (2006).

\bibitem{s20}
I. Wilson-Rae et al., Phys. Rev. Lett. {\bf 99}, 093901 (2007); T.
J. Kippenberg and K. J. Vahala, Opt. Express {\bf 15}, 17172
(2007).

\bibitem{s21}
F. Marquardt et al., Phys. Rev. Lett. {\bf 99}, 093902 (2007);
F.Marquardt et al., J. Mod. Opt. 55, 3329 (2008).

\bibitem{s22}
C. Genes et al., Phys. Rev. A {\bf 77}, 033804 (2008); M. Grajcar
et al., Phys. Rev. B {\bf 78}, 035406 (2008).

\bibitem{s23}
P. Rabl, C. Genes, K. Hammerer and M. Aspelmeyer, arXiv:
0903.1637; S. Gr¡§oblacher, K. Hammerer, M. R. Vanner, M.
Aspelmeyer, arXiv: 0903.5293.

\bibitem{s24}
M. D. LaHaye, O. Buu, B. Camarota and K. C. Schwab, Science {\bf
304}, 74 (2004).

\bibitem{s25}
F. Diedrich et al., Phys. Rev. Lett. {\bf 62}, 403 (1989); C.
Monroe et al., Phys. Rev. Lett. {\bf 75}, 4011 (1995).

\bibitem{s26}
S. E. Hamann et al., Phys. Rev. Lett. {\bf 80}, 4149 (1998).

\bibitem{s27}
X. Y. Wu, B. J. Zhang, H, B. Li, X. J. Liu, J. W. Li and Y. Q.
Guo, Int. J. Theor. Phys. {\bf 48}, 2027 (2009).

\bibitem{s28}
T.Walker, D.Sesko and C.Wieman, PRL, {\bf 64}, 408 (1990).

\bibitem{s29}
T.Loftus, J.R.Bochinski and T.W.Mossberg, Phys.Rev A {\bf 63},
053401 (2001).

\bibitem{s30}
V.Bagnato, L.Marcassa, M.Oria, G.Surdutovich, R.Vitlina and
S.Zilio, Phys. Rev.A {\bf 48}, 3771 (1993).

\bibitem{s31}
V.Bagnato, L.Marcassa, M.Oria, G.Surdutovich and S.Zilio, Laser
Physics {\bf 2}, 172 (1992).

\bibitem{s32}
A. D. O¡¯Connell, M. Hofheinz, M. Ansmann, Radoslaw C. Bialczak,
M. Lenander, Erik Lucero, M. Neeley, D. Sank, H. Wang, M. Weides,
J. Wenner, John M. Martinis and A. N. Cleland, Nature {\bf 464},
697 (2010).



\end{thebibliography}
\end{document}